\documentclass[fleqn,twoside]{article}
\usepackage{espcrc2}

\usepackage{graphicx}
\usepackage[figuresright]{rotating}

\newcommand{\AmS}{{\protect\the\textfont2
  A\kern-.1667em\lower.5ex\hbox{M}\kern-.125emS}}

\title{Neutrino telescopes as a probe of active and sterile neutrino mixings}

\author{Zhi-zhong Xing\address[MCSD]{Institute of High Energy Physics,
P.O. Box 918, Beijing 100049, China}}

\begin{document}

\begin{abstract}
If the ultrahigh-energy (UHE) neutrino fluxes produced from a
distant astrophysical source can be measured at a km$^3$-size
neutrino telescope, they will provide a promising way to help
determine the flavor mixing pattern of three active neutrinos.
Considering the conventional UHE neutrino source with the flavor
ratio $\phi^{}_e : \phi^{}_\mu : \phi^{}_\tau = 1 : 2 : 0$, I show
that $\phi^{\rm D}_e : \phi^{\rm D}_\mu : \phi^{\rm D}_\tau = (1
-2 \Delta) : (1 +\Delta) : (1 +\Delta)$ holds at the detector of a
neutrino telescope, where $\Delta$ characterizes the effect of
$\mu$-$\tau$ symmetry breaking (i.e., $\theta^{}_{13} \neq 0$ and
$\theta^{}_{23} \neq \pi/4$). Current experimental data yield
$-0.1 \leq \Delta \leq +0.1$. It is also possible to probe
$\Delta$ by detecting the $\overline{\nu}^{}_e$ flux of
$E^{}_{\overline{\nu}^{}_e} \approx 6.3 ~ {\rm PeV}$ via the
Glashow resonance channel $\overline{\nu}^{}_e e \rightarrow W^-
\rightarrow ~ {\rm anything}$. Finally, I give some brief comments
on the possibility to constrain the mixing between active and
sterile neutrinos by using the UHE neutrino telescopes.
\end{abstract}

\maketitle

\section{Introduction}

The solar \cite{SNO}, atmospheric \cite{SK}, reactor \cite{KM} and
accelerator \cite{K2K} neutrino experiments have provided
convincing evidence for the existence of neutrino oscillations and
opened a window to new physics beyond the standard model. The
neutrino mixing is described by a unitary matrix $V$,
\begin{equation}
\left ( \matrix{\nu^{}_e \cr \nu^{}_\mu \cr \nu^{}_\tau \cr}
\right ) \; =\; \left ( \matrix{ V^{}_{e1} & V^{}_{e2} & V^{}_{e3}
\cr V^{}_{\mu 1} & V^{}_{\mu 2} & V^{}_{\mu 3} \cr V^{}_{\tau 1} &
V^{}_{\tau 2} & V^{}_{\tau 3} \cr} \right ) \left ( \matrix{
\nu^{}_1 \cr \nu^{}_2 \cr \nu^{}_3 \cr} \right ) \; .
\end{equation}
In the ``standard" parametrization of $V$ \cite{PDG}, one defines
$V^{}_{e2} = \sin\theta^{}_{12} \cos\theta^{}_{13}$, $V^{}_{e3} =
\sin\theta^{}_{13} e^{-i\delta}$ and $V^{}_{\mu 3} =
\sin\theta^{}_{23} \cos\theta^{}_{13}$. Here I have omitted the
Majorana CP-violating phases from $V$, because they are irrelevant
to the properties of neutrino oscillations to be discussed. A
global analysis of current experimental data (see, e.g., Ref.
\cite{Vissani}) points to $\theta^{}_{13} = 0$ and $\theta^{}_{23}
= \pi/4$, which motivate a number of authors to consider the
$\mu$-$\tau$ permutation symmetry for model building
\cite{Symmetry}.

The main purpose of my talk is to investigate how the effect of
$\mu$-$\tau$ symmetry breaking can show up at a neutrino
telescope. I anticipate that IceCube \cite{Ice} and other
second-generation neutrino telescopes \cite{Water} are able to
detect the fluxes of ultrahigh-energy (UHE) $\nu^{}_e$
($\overline{\nu}^{}_e$), $\nu^{}_\mu$ ($\overline{\nu}^{}_\mu$)
and $\nu^{}_\tau$ ($\overline{\nu}^{}_\tau$) neutrinos generated
from very distant astrophysical sources. For most of the
currently-envisaged sources of UHE neutrinos \cite{R}, a general
and canonical expectation is that the initial neutrino fluxes are
produced via the decay of pions created from $pp$ or $p\gamma$
collisions and their flavor content can be expressed as
\begin{equation}
\left \{\phi^{}_e ~,~ \phi^{}_\mu ~,~ \phi^{}_\tau \right \} \; =
\; \left \{ \frac{1}{3} ~,~ \frac{2}{3} ~,~ 0 \right \} \phi^{}_0
\; ,
\end{equation}
where $\phi^{}_\alpha$ (for $\alpha = e, \mu, \tau$) denotes the
sum of $\nu^{}_\alpha$ and $\overline{\nu}^{}_\alpha$ fluxes, and
$\phi^{}_0 = \phi^{}_e + \phi^{}_\mu + \phi^{}_\tau$ is the total
flux of neutrinos and antineutrinos of all flavors. Due to
neutrino oscillations, the flavor composition of such cosmic
neutrino fluxes to be measured at the detector of a neutrino
telescope has been expected to be \cite{Pakvasa}
\begin{equation}
\left \{\phi^{\rm D}_e ,~ \phi^{\rm D}_\mu ,~ \phi^{\rm D}_\tau
\right \} \; = \; \left \{ \frac{1}{3} ~,~ \frac{1}{3} ~,~
\frac{1}{3} \right \} \phi^{}_0 \; .
\end{equation}
However, it is worth remarking that this naive expectation is only
true in the limit of $\mu$-$\tau$ symmetry (or equivalently,
$\theta^{}_{13} = 0$ and $\theta^{}_{23} = \pi/4$). Starting from
the hypothesis given in Eq. (2) and allowing for the slight
breaking of $\mu$-$\tau$ symmetry, I am going to show that
\begin{eqnarray}
\phi^{\rm D}_e : \phi^{\rm D}_\mu : \phi^{\rm D}_\tau  = \left (1
-2 \Delta \right ) : \left (1 +\Delta \right ) : \left (1 +\Delta
\right ) \;
\end{eqnarray}
holds to an excellent degree of accuracy, where $\Delta$
characterizes the effect of $\mu$-$\tau$ symmetry breaking (i.e.,
the combined effect of $\theta^{}_{13} \neq 0$ and $\theta^{}_{23}
\neq \pi/4$) \cite{XingZZ}. I obtain $-0.1 \leq \Delta \leq +0.1$
from the present neutrino oscillation data. I find that it is also
possible to probe $\Delta$ by detecting the $\overline{\nu}^{}_e$
flux of $E^{}_{\overline{\nu}^{}_e} \approx 6.3 ~ {\rm PeV}$ via
the well-known Glashow resonance (GR) \cite{Glashow} channel
$\overline{\nu}^{}_e e \rightarrow W^- \rightarrow ~ {\rm
anything}$ at a neutrino telescope. Finally, I will give some
comments on the possibility to constrain the mixing between active
and sterile neutrinos by using neutrino telescopes.

\section{Signals of $\mu$-$\tau$ symmetry breaking}

Let me define $\phi^{(\rm D)}_\alpha \equiv \phi^{(\rm
D)}_{\nu^{}_\alpha} + \phi^{(\rm D)}_{\overline{\nu}^{}_\alpha}$
(for $\alpha = e, \mu, \tau$) throughout this paper, where
$\phi^{(\rm D)}_{\nu^{}_\alpha}$ and $\phi^{(\rm
D)}_{\overline{\nu}^{}_\alpha}$ denote the $\nu^{}_\alpha$ and
$\overline{\nu}^{}_\alpha$ fluxes, respectively. As for the UHE
neutrino fluxes produced from the pion-muon decay chain with
$\phi^{}_{\nu^{}_\tau} = \phi^{}_{\overline{\nu}^{}_\tau} = 0$,
the relationship between $\phi^{}_{\nu^{}_\alpha}$ (or
$\phi^{}_{\overline{\nu}^{}_\alpha}$) and $\phi^{\rm
D}_{\nu^{}_\alpha}$ (or $\phi^{\rm D}_{\overline{\nu}^{}_\alpha}$)
is given by $\phi^{\rm D}_{\nu^{}_\alpha} = \phi^{}_{\nu^{}_e}
P^{}_{e\alpha} + \phi^{}_{\nu^{}_\mu} P^{}_{\mu\alpha}$ or
$\phi^{\rm D}_{\overline{\nu}^{}_\alpha} =
\phi^{}_{\overline{\nu}^{}_e} \bar{P}^{}_{e\alpha} +
\phi^{}_{\overline{\nu}^{}_\mu} \bar{P}^{}_{\mu\alpha}$, in which
$P^{}_{\beta\alpha}$ and $\bar{P}^{}_{\beta\alpha}$ (for $\alpha =
e, \mu, \tau$ and $\beta = e$ or $\mu$) stand respectively for the
oscillation probabilities $P (\nu^{}_\beta \rightarrow
\nu^{}_\alpha)$ and $P (\overline{\nu}^{}_\beta \rightarrow
\overline{\nu}^{}_\alpha)$. Because the Galactic distances far
exceed the observed neutrino oscillation lengths,
$P^{}_{\beta\alpha}$ and $\bar{P}^{}_{\beta\alpha}$ are actually
averaged over many oscillations. Then I obtain
$\bar{P}^{}_{\beta\alpha} = P^{}_{\beta\alpha}$ and
\begin{equation}
P^{}_{\beta\alpha} \; = \; \sum^3_{i=1} |V^{}_{\alpha i}|^2
|V^{}_{\beta i}|^2 \; ,
\end{equation}
where $V^{}_{\alpha i}$ and $V^{}_{\beta i}$ (for $\alpha, \beta =
e, \mu, \tau$ and $i = 1, 2, 3$) denote the matrix elements of $V$
defined in Eq. (1). The relationship between $\phi^{}_\alpha$ and
$\phi^{\rm D}_\alpha$ turns out to be
\begin{equation}
\phi^{\rm D}_\alpha \; = \; \phi^{}_e P^{}_{e\alpha} + \phi^{}_\mu
P^{}_{\mu\alpha} \; .
\end{equation}
To be explicit, I have
\begin{eqnarray}
\phi^{\rm D}_e & = & \frac{\phi^{}_0}{3} \left (P^{}_{ee} + 2
P^{}_{\mu e} \right ) \; , \nonumber \\
\phi^{\rm D}_\mu & = & \frac{\phi^{}_0}{3} \left (P^{}_{e\mu} + 2
P^{}_{\mu \mu} \right ) \; , \nonumber \\
\phi^{\rm D}_\tau & = & \frac{\phi^{}_0}{3} \left (P^{}_{e\tau} +
2P^{}_{\mu \tau} \right ) \; .
\end{eqnarray}
It is then possible to evaluate the relative sizes of $\phi^{\rm
D}_e$, $\phi^{\rm D}_\mu$ and $\phi^{\rm D}_\tau$ by using Eqs.
(1), (5) and (7).

In order to clearly show the effect of $\mu$-$\tau$ symmetry
breaking on the neutrino fluxes to be detected at neutrino
telescopes, I define
\begin{equation}
\varepsilon \; \equiv \; \theta^{}_{23} - \frac{\pi}{4} \; ,
~~~~~~~ (|\varepsilon| \ll 1) \; .
\end{equation}
Namely, $\varepsilon$ measures the slight departure of
$\theta^{}_{23}$ from $\pi/4$. Using small $\theta^{}_{13}$ and
$\varepsilon$, I express $|V^{}_{\alpha i}|^2$ (for $\alpha =e,
\mu, \tau$ and $i=1,2,3$) as follows:
\begin{eqnarray}
\left [ |V^{}_{\alpha i}|^2 \right ] & = & \frac{1}{2} A +
\varepsilon B + \frac{1}{2} \left ( \theta^{}_{13} \sin
2\theta^{}_{12} \cos\delta \right ) C
\nonumber \\
&& + {\cal O}(\varepsilon^2) + {\cal O}(\theta^2_{13}) \; .
\end{eqnarray}
where
\begin{eqnarray}
A & = & \left [ \matrix{ 2\cos^2\theta^{}_{12} &
2\sin^2\theta^{}_{12} & 0 \cr \sin^2\theta^{}_{12} &
\cos^2\theta^{}_{12} & 1 \cr \sin^2\theta^{}_{12} &
\cos^2\theta^{}_{12} & 1 \cr} \right ] \; ,
\nonumber \\
B & = & \left [ \matrix{ 0 & 0 & 0 \cr -\sin^2\theta^{}_{12} &
-\cos^2\theta^{}_{12} & 1 \cr \sin^2\theta^{}_{12} &
\cos^2\theta^{}_{12} & -1 \cr} \right ] \; ,
\nonumber \\
C & = & \left [ \matrix{ 0 & 0 & 0 \cr 1 & -1 & 0 \cr -1 & 1 & 0
\cr} \right ] \; . \nonumber
\end{eqnarray}
Eqs. (5) and (9) allow me to calculate $P^{}_{\beta\alpha}$:
\begin{eqnarray}
P^{}_{ee} + 2 P^{}_{\mu e} & = & 1 + \frac{\theta^{}_{13}}{2} \sin
4\theta^{}_{12} \cos\delta
\nonumber \\
&& - \varepsilon \sin^2 2\theta^{}_{12} \; ,
\nonumber \\
P^{}_{e\mu} + 2 P^{}_{\mu \mu} & = & 1 - \frac{\theta^{}_{13}}{4}
\sin 4\theta^{}_{12} \cos\delta
\nonumber \\
&& + \frac{\varepsilon}{2} \sin^2 2\theta^{}_{12} \; ,
\nonumber \\
P^{}_{e\tau} + 2 P^{}_{\mu \tau} & = & 1 -
\frac{\theta^{}_{13}}{4} \sin 4\theta^{}_{12} \cos\delta
\nonumber \\
&& + \frac{\varepsilon}{2} \sin^2 2\theta^{}_{12} \; ,
\end{eqnarray}
where the terms of ${\cal O} (\varepsilon^2)$ and ${\cal
O}(\theta^2_{13})$ are omitted. Substituting Eq. (10) into Eq.
(7), I get \cite{XingZZ}
\begin{eqnarray}
\phi^{\rm D}_e & = & \frac{\phi^{}_0}{3} \left ( 1 - 2\Delta
\right )
\; , \nonumber \\
\phi^{\rm D}_\mu & = & \frac{\phi^{}_0}{3} \left ( 1 + \Delta
\right )
\; , \nonumber \\
\phi^{\rm D}_\tau & = & \frac{\phi^{}_0}{3} \left ( 1 + \Delta
\right ) \; ,
\end{eqnarray}
where
\begin{equation}
\Delta \; = \; \frac{1}{4} \left ( 2\varepsilon \sin^2
2\theta^{}_{12} - \theta^{}_{13} \sin 4\theta^{}_{12} \cos\delta
\right ) \; .
\end{equation}
Eq. (4) is therefore proved by Eq. (11). One can see that
$\phi^{\rm D}_e + \phi^{\rm D}_\mu + \phi^{\rm D}_\tau =
\phi^{}_0$ holds. Some discussions are in order.

(1) The small parameter $\Delta$ characterizes the overall effect
of $\mu$-$\tau$ symmetry breaking. Allowing $\delta$ to vary
between $0$ and $\pi$, I obtain the lower and upper bounds of
$\Delta$ for given values of $\theta^{}_{12}$ ($< \pi/4$),
$\theta^{}_{13}$ and $\varepsilon$: $-\Delta^{}_{\rm bound} \leq
\Delta \leq +\Delta^{}_{\rm bound}$, where
\begin{equation}
\Delta^{}_{\rm bound} = \frac{1}{4} \left ( 2|\varepsilon| \sin^2
2\theta^{}_{12} + \theta^{}_{13} \sin 4\theta^{}_{12}\right ) \; .
\end{equation}
It is obvious that $\Delta = -\Delta^{}_{\rm bound}$ when
$\varepsilon <0$ and $\delta =0$, and $\Delta = +\Delta^{}_{\rm
bound}$ when $\varepsilon >0$ and $\delta =\pi$. A global analysis
of current neutrino oscillation data \cite{Vissani} indicates
$30^\circ < \theta^{}_{12} < 38^\circ$, $\theta^{}_{13} <
10^\circ$ ($\approx 0.17$) and $|\varepsilon| < 9^\circ$ ($\approx
0.16$) at the $99\%$ confidence level, but the CP-violating phase
$\delta$ is entirely unrestricted. Using these constraints, I
analyze the allowed range of $\Delta$ and its dependence on
$\delta$. The maximal value of $\Delta^{}_{\rm bound}$ (i.e.,
$\Delta^{}_{\rm bound} \approx 0.098$) appears when
$|\varepsilon|$ and $\theta^{}_{13}$ approach their respective
upper limits and $\theta^{}_{12} \approx 33^\circ$ holds
\cite{XingZZ}. $\Delta^{}_{\rm bound}$ is not very sensitive to
the variation of $\theta^{}_{12}$ in its allowed region.

If $\theta^{}_{13} =0$ holds, $\Delta^{}_{\rm bound} =
0.5|\varepsilon|\sin^2 2\theta^{}_{12} <0.074$ when
$\theta^{}_{12}$ approaches its upper limit. If $\varepsilon =0$
(i.e., $\theta^{}_{23} =\pi/4$) holds, I obtain $\Delta^{}_{\rm
bound} = 0.25 \theta^{}_{13} \sin 4\theta^{}_{12} <0.038$ when
$\theta^{}_{12}$ approaches its lower limit. Thus $\Delta^{}_{\rm
bound}$ is more sensitive to the deviation of $\theta^{}_{23}$
from $\pi/4$.

(2) Of course, $\Delta =0$ exactly holds when $\theta^{}_{13} =
\varepsilon =0$ is taken. Because the sign of $\varepsilon$ and
the range of $\delta$ are both unknown, we are now unable to rule
out the nontrivial possibility $\Delta \approx 0$ in the presence
of $\theta^{}_{13} \neq 0$ and $\varepsilon \neq 0$. In other
words, $\Delta$ may be vanishing or extremely small if its two
leading terms cancel each other. It is easy to arrive at $\Delta
\approx 0$ from Eq. (12), if the condition
\begin{equation}
\frac{\varepsilon}{\theta^{}_{13}} \; =\; \cot 2\theta^{}_{12}
\cos\delta \;
\end{equation}
is satisfied. Because of $|\cos\delta| \leq 1$, Eq. (14) imposes a
strong constraint on the magnitude of
$\varepsilon/\theta^{}_{13}$. The dependence of
$\varepsilon/\theta^{}_{13}$ on $\delta$ is illustrated in Ref.
\cite{XingZZ}, where $\theta^{}_{12}$ varies in its allowed range.
I find that $|\varepsilon|/\theta^{}_{13} < 0.6$ is necessary to
hold, such that a large cancellation between two leading terms of
$\Delta$ is possible to take place.

The implication of the above result on UHE neutrino telescopes is
two-fold. On the one hand, an observable signal of $\Delta \neq 0$
at a neutrino telescope implies the existence of significant
$\mu$-$\tau$ symmetry breaking. If a signal of $\Delta \neq 0$
does not show up at a neutrino telescope, on the other hand, one
cannot conclude that the $\mu$-$\tau$ symmetry is an exact or
almost exact symmetry. It is therefore meaningful to consider the
complementarity between neutrino telescopes and terrestrial
neutrino oscillation experiments \cite{Winter}, in order to
finally pin down the parameters of neutrino mixing and leptonic CP
violation.

(3) To illustrate, I define the flux ratios
\begin{eqnarray}
R^{}_e & \equiv & \frac{\phi^{\rm D}_e}{\phi^{\rm D}_\mu +
\phi^{\rm D}_\tau} \; ,
\nonumber \\
R^{}_\mu & \equiv & \frac{\phi^{\rm D}_\mu}{\phi^{\rm D}_\tau +
\phi^{\rm D}_e} \; ,
\nonumber \\
R^{}_\tau & \equiv & \frac{\phi^{\rm D}_\tau}{\phi^{\rm D}_e +
\phi^{\rm D}_\mu} \; ,
\end{eqnarray}
which may serve as the {\it working} observables at neutrino
telescopes \cite{XZ}. At least, $R^{}_\mu$ can be extracted from
the ratio of muon tracks to showers at IceCube \cite{Ice}, even if
those electron and tau events cannot be disentangled. Taking
account of Eq. (11), I approximately obtain
\begin{eqnarray}
R^{}_e & \approx & \frac{1}{2} ~ - ~ \frac{3}{2} \Delta \; ,
\nonumber \\
R^{}_\mu & \approx & \frac{1}{2} ~ + ~ \frac{3}{4} \Delta \; ,
\nonumber \\
R^{}_\tau & \approx & \frac{1}{2} ~ + ~ \frac{3}{4} \Delta \; .
\end{eqnarray}
It turns out that $R^{}_e$ is most sensitive to the effect of
$\mu$-$\tau$ symmetry breaking.

Due to $\phi^{\rm D}_\mu = \phi^{\rm D}_\tau$ shown in Eq. (11),
$R^{}_\mu = R^{}_\tau$ holds no matter whether $\Delta$ vanishes
or not. This observation implies that the ``$\mu$-$\tau$" symmetry
between $R^{}_\mu$ and $R^{}_\tau$ is actually insensitive to the
breaking of $\mu$-$\tau$ symmetry in the neutrino mass matrix. If
both $R^{}_e$ and $R^{}_\mu$ are measured, one can then extract
$\Delta$ from their difference:
\begin{equation}
R^{}_\mu - R^{}_e \; =\; \frac{9}{4} \Delta \; .
\end{equation}
Taking $\Delta = \Delta^{}_{\rm bound} \approx 0.1$, we get
$R^{}_\mu - R^{}_e \leq 0.22$.

\section{On the Glashow resonance}

I proceed to discuss the possibility to probe the breaking of
$\mu$-$\tau$ symmetry by detecting the $\overline{\nu}^{}_e$ flux
from distant astrophysical sources through the so-called Glashow
resonance (GR) channel $\overline{\nu}^{}_e e \rightarrow W^-
\rightarrow ~ {\rm anything}$ \cite{Glashow}. The latter can take
place over a very narrow energy interval around the
$\overline{\nu}^{}_e$ energy $E^{\rm GR}_{\overline{\nu}^{}_e}
\approx M^2_W/2m^{}_e \approx 6.3 ~ {\rm PeV}$. A neutrino
telescope may measure both the GR-mediated $\overline{\nu}^{}_e$
events ($N^{\rm GR}_{\overline{\nu}^{}_e}$) and the $\nu^{}_\mu +
\overline{\nu}^{}_\mu$ events of charged-current (CC) interactions
($N^{\rm CC}_{\nu^{}_\mu + \overline{\nu}^{}_\mu}$) in the
vicinity of $E^{\rm GR}_{\overline{\nu}^{}_e}$. Their ratio,
defined as $R^{}_{\rm RG} \equiv N^{\rm
GR}_{\overline{\nu}^{}_e}/N^{\rm CC}_{\nu^{}_\mu +
\overline{\nu}^{}_\mu}$, can be related to the ratio of
$\overline{\nu}^{}_e$'s to $\nu^{}_\mu$'s and
$\overline{\nu}^{}_\mu$'s entering the detector,
\begin{equation}
R^{}_0 \; \equiv \; \frac{\phi^{\rm
D}_{\overline{\nu}^{}_e}}{\phi^{\rm D}_{\nu^{}_\mu} + \phi^{\rm
D}_{\overline{\nu}^{}_\mu}} \; .
\end{equation}
Note that $\phi^{\rm D}_{\overline{\nu}^{}_e}$, $\phi^{\rm
D}_{\nu^{}_\mu}$ and $\phi^{\rm D}_{\overline{\nu}^{}_\mu}$ stand
respectively for the fluxes of $\overline{\nu}^{}_e$'s,
$\nu^{}_\mu$'s and $\overline{\nu}^{}_\mu$'s before the RG and CC
interactions occur at the detector. In a recent paper \cite{BG},
$R^{}_{\rm GR} = a R^{}_0$ with $a \approx 30.5$ has been obtained
by considering the muon events with contained vertices
\cite{Beacom} in a water- or ice-based detector. An accurate
calculation of $a$ is crucial for a specific neutrino telescope to
detect the GR reaction rate, but it is beyond the scope of this
talk. Here I only concentrate on the possible effect of
$\mu$-$\tau$ symmetry breaking on $R^{}_0$.

Provided the initial neutrino fluxes are produced via the decay of
$\pi^+$'s and $\pi^-$'s created from high-energy $pp$ collisions,
their flavor composition can be expressed in a more detailed way
as
\begin{eqnarray}
\left \{\phi^{}_{\nu^{}_e} ,~ \phi^{}_{\nu^{}_\mu} ,~
\phi^{}_{\nu^{}_\tau} \right \} & = & \left \{ \frac{1}{6} ~,~
\frac{1}{3} ~,~ 0 \right \} \phi^{}_0 \; ,
\nonumber \\
\left \{\phi^{}_{\overline{\nu}^{}_e} ,~
\phi^{}_{\overline{\nu}^{}_\mu} ,~
\phi^{}_{\overline{\nu}^{}_\tau} \right \} & = & \left \{
\frac{1}{6} ~,~ \frac{1}{3} ~,~ 0 \right \} \phi^{}_0 \; .
\end{eqnarray}
In comparison, the flavor content of UHE neutrino fluxes produced
from $p\gamma$ collisions reads
\begin{eqnarray}
\left \{\phi^{}_{\nu^{}_e} ,~ \phi^{}_{\nu^{}_\mu} ,~
\phi^{}_{\nu^{}_\tau} \right \} & = & \left \{ \frac{1}{3} ~,~
\frac{1}{3} ~,~ 0 \right \} \phi^{}_0 \; ,
\nonumber \\
\left \{\phi^{}_{\overline{\nu}^{}_e} ,~
\phi^{}_{\overline{\nu}^{}_\mu} ,~
\phi^{}_{\overline{\nu}^{}_\tau} \right \} & = & \left \{ 0 ~,~
\frac{1}{3} ~,~ 0 \right \} \phi^{}_0 \; .
\end{eqnarray}
For either Eq. (19) or Eq. (20), the sum of
$\phi^{}_{\nu^{}_\alpha}$ and $\phi^{}_{\overline{\nu}^{}_\alpha}$
is consistent with $\phi^{}_\alpha$ in Eq. (2).

Due to neutrino oscillations, the $\overline{\nu}^{}_e$ flux at
the detector of a neutrino telescope is given by $\phi^{\rm
D}_{\overline{\nu}^{}_e} = \phi^{}_{\overline{\nu}^{}_e}
\bar{P}^{}_{ee} + \phi^{}_{\overline{\nu}^{}_\mu} \bar{P}^{}_{\mu
e}$. With the help of Eqs. (5), (9), (19) and (20), I explicitly
obtain
\begin{eqnarray}
\phi^{\rm D}_{\overline{\nu}^{}_e} (pp) ~ & = & ~
\frac{\phi^{}_0}{6} \left (1 - 2 \Delta \right ) \; ,
\nonumber \\
\phi^{\rm D}_{\overline{\nu}^{}_e} (p\gamma) ~ & = & ~
\frac{\phi^{}_0}{12} \left (\sin^2 2\theta^{}_{12} - 4 \Delta
\right ) \; .
\end{eqnarray}
The sum of $\phi^{\rm D}_{\nu^{}_\mu}$ and $\phi^{\rm
D}_{\overline{\nu}^{}_\mu}$, which is defined as $\phi^{\rm
D}_\mu$, has been given in Eq. (11). It is then straightforward to
calculate $R^{}_0$ by using Eq. (18) for two different
astrophysical sources:
\begin{eqnarray}
R^{}_0(pp) ~ & \approx & ~ \frac{1}{2} ~ - ~\frac{3}{2}\Delta \; ,
\nonumber \\
R^{}_0(p\gamma) ~ & \approx & ~ \frac{\sin^2 2\theta^{}_{12}}{4} ~
- ~ \frac{4 + \sin^2 2\theta^{}_{12}}{4} \Delta \; .
\end{eqnarray}
This result indicates that the dependence of $R^{}_0(pp)$ on
$\theta^{}_{12}$ is hidden in $\Delta$ and suppressed by the
smallness of $\theta^{}_{13}$ and $\varepsilon$. In addition, the
deviation of $R^{}_0(pp)$ from $1/2$ can be as large as
$1.5\Delta^{}_{\rm bound} \approx 0.15$. It is obvious that the
ratio $R^{}_0(p\gamma)$ is very sensitive to the value of $\sin^2
2\theta^{}_{12}$. A measurement of $R^{}_0 (p\gamma)$ at IceCube
and other second-generation neutrino telescopes may therefore
probe the mixing angle $\theta^{}_{12}$ \cite{BG}. Indeed, the
dominant production mechanism for ultrahigh-energy neutrinos at
Active Galactic Nuclei (AGNs) and Gamma Ray Bursts (GRBs) is
expected to be the $p\gamma$ process in a tenuous or
radiation-dominated environment \cite{Berezinsky}. If this
expectation is true, the observation of $R^{}_0(p\gamma)$ may also
provide us with useful information on the breaking of $\mu$-$\tau$
symmetry.

\section{Comments on sterile neutrinos}

Today we are not well motivated to consider the existence of very
light sterile neutrinos, which may take part in the oscillations
of active neutrinos and change the conventional interpretation of
current experimental results \cite{Vissani}. In particular, it is
hard to simultaneously interpret the LSND anomaly \cite{LSND} and
other convincing neutrino oscillation data by introducing one or
two sterile neutrinos. The mixing between sterile and active
neutrinos has to be sufficiently suppressed; otherwise, it might
bring sterile neutrinos in equilibrium with active neutrinos
before neutrino decoupling --- the resultant excess in energy
dependence would endanger the Big Bang nucleosynthesis of light
elements \cite{Barbieri}. If the mass-squared differences between
active and sterile neutrinos are of ${\cal O}(10^{-11}) ~ {\rm
eV}^2$ or smaller, however, neutrino oscillations will not produce
and maintain a significant sterile neutrino population. This case
has been considered in Ref. \cite{Maalampi} with a conclusion that
the UHE neutrinos may offer a unique opportunity to probe neutrino
oscillations in the mass-squared range $10^{-16} ~ {\rm eV}^2 \leq
\Delta m^2 \leq 10^{-11} ~ {\rm eV}^2$, a region which is not
accessible by any other means.

The possible effects of sterile neutrinos on the UHE neutrino
fluxes have been discussed in the literature (see, e.g., Refs.
\cite{Maalampi} and \cite{Yasuda}). For simplicity, I do not
elaborate on them in this talk. I only emphasize that some of such
discussions are already out of date, because the experimental
constraints on the mixing between sterile and active neutrinos
have become more stringent than before. Whether the light sterile
neutrinos exist or not remains an open question.

\section{Comments on the ratio $\phi^{}_e : \phi^{}_\mu :
\phi^{}_\tau$}

What I have so far considered is the canonical or conventional
astrophysical source, from which the UHE neutrino flux results
from the pion decays and thus has the flavor composition
$\phi^{}_e : \phi^{}_\mu : \phi^{}_\tau = 1: 2 : 0$. In reality,
however, this simple flavor content could somehow by contaminated
for certain reasons (e.g., a small amount of $\nu^{}_e$,
$\nu^{}_\mu$ and $\nu^{}_\tau$ and their antiparticles might come
from the decays of heavier hadrons produced by $pp$ and $p\gamma$
collisions. Following a phenomenological approach, Zhou and I
\cite{XZ} proposed a generic parametrization of the initial flavor
composition of an UHE neutrino flux:
\begin{equation}
\left ( \matrix{ \phi^{}_e \cr \phi^{}_\mu \cr \phi^{}_\tau \cr}
\right ) \; = \; \left ( \matrix{ \sin^2 \xi \cos^2 \zeta \cr
\cos^2 \xi \cos^2 \zeta \cr \sin^2 \zeta \cr} \right ) \phi^{}_0
\; ,
\end{equation}
where $\xi \in [0, \pi/2]$ and $\zeta \in [0, \pi/2]$. Then the
conventional picture, as shown in Eq. (2), corresponds to $\zeta =
0$ and $\tan\xi = 1/\sqrt{2}$ (or $\xi \approx 35.3^\circ$) in our
parametrization. It turns out that any small departure of $\zeta$
from zero will measure the existence of cosmic $\nu^{}_\tau$ and
$\overline{\nu}^{}_\tau$ neutrinos, which could come from the
decays of $D^{}_s$ and $B\overline{B}$ mesons produced at the
source \cite{Pakvasa}. On the other hand, any small deviation of
$\tan^2\xi$ from $1/2$ will imply that the pure pion-decay
mechanism for the UHE neutrino production has to be modified.

After defining three neutrino flux ratios $R^{}_\alpha$ (see Eq.
(15) for $\alpha = e, \mu, \tau$) as our working observables at a
neutrino telescope, we have shown that the source parameters $\xi$
and $\zeta$ can in principle be determined by the measurement of
two independent $R^{}_\alpha$ and with the help of accurate
neutrino oscillation data \cite{XZ}. We have also examined the
dependence of $R^{}_\alpha$ upon the smallest neutrino mixing
angle $\theta^{}_{13}$ and upon the Dirac CP-violating phase
$\delta$. Our numerical examples indicate that it is promising to
determine or (at least) constrain the initial flavor content of
UHE neutrino fluxes with the second-generation neutrino
telescopes.

\section{Concluding remarks}

I have discussed why and how the second-generation neutrino
telescopes can serve as a striking probe of broken $\mu$-$\tau$
symmetry. Based on the conventional mechanism for UHE neutrino
production at a distant astrophysical source and the standard
picture of neutrino oscillations, I have shown that the flavor
composition of cosmic neutrino fluxes at a terrestrial detector
may deviate from the naive expectation $\phi^{\rm D}_e : \phi^{\rm
D}_\mu : \phi^{\rm D}_\tau = 1 : 1 : 1$. Instead, $\phi^{\rm D}_e
: \phi^{\rm D}_\mu : \phi^{\rm D}_\tau = (1 -2 \Delta) : (1
+\Delta) : (1 +\Delta)$ holds, where $\Delta$ characterizes the
effect of $\mu$-$\tau$ symmetry breaking. The latter is actually a
reflection of $\theta^{}_{13} \neq 0$ and $\theta^{}_{23} \neq
\pi/4$ in the $3\times 3$ neutrino mixing matrix. I have examined
the sensitivity of $\Delta$ to the deviation of $\theta^{}_{13}$
from zero and to the departure of $\theta^{}_{23}$ from $\pi/4$,
and obtained $-0.1 \leq \Delta \leq +0.1$ from current data. I
find that it is also possible to probe the breaking of
$\mu$-$\tau$ symmetry by detecting the $\overline{\nu}^{}_e$ flux
of $E^{}_{\overline{\nu}^{}_e} \approx 6.3 ~ {\rm PeV}$ via the
Glashow resonance channel $\overline{\nu}^{}_e e \rightarrow W^-
\rightarrow ~ {\rm anything}$.

This work, different from the previous ones (see Refs.
\cite{Winter,XZ,BG,Serpico}) in studying how to determine or
constrain one or two of three neutrino mixing angles and the Dirac
CP-violating phase with neutrino telescopes, reveals the combined
effect of $\theta^{}_{13} \neq 0$, $\theta^{}_{23} \neq \pi/4$ and
$\delta \neq \pi/2$ which can show up at the detector. Even if
$\Delta \neq 0$ is established from the measurement of UHE
neutrino fluxes, the understanding of this $\mu$-$\tau$ symmetry
breaking signal requires more precise information about
$\theta^{}_{13}$, $\theta^{}_{23}$ and $\delta$. Hence it makes
sense to look at the complementary roles played by neutrino
telescopes and terrestrial neutrino oscillation experiments (e.g.,
the reactor experiments to pin down the smallest neutrino mixing
angle $\theta^{}_{13}$ and the neutrino factories or superbeam
facilities to measure the CP-violating phase $\delta$) in the era
of precision measurements.

The feasibility of the above idea depends on the assumption that
we have correctly understood the production mechanism of cosmic
neutrinos from a distant astrophysical source (i.e., via $pp$ and
$p\gamma$ collisions) with little uncertainties. It is also
dependent upon the assumption that the error bars associated with
the measurement of relevant neutrino fluxes or their ratios are
much smaller than $\Delta$. The latter is certainly a challenge to
the sensitivity or precision of IceCube and other neutrino
telescopes under construction or under consideration, unless the
effect of $\mu$-$\tau$ symmetry breaking is unexpectedly large.
Nevertheless, any constraint on $\Delta$ to be obtained from
neutrino telescopes will be greatly useful in diagnosing the
astrophysical sources and in understanding the properties of
neutrinos themselves. Much more effort is therefore needed in this
direction.

Finally, I would like to thank Y.Q. Ma and other organizers for
kind invitation and warm hospitality. The symposium is as
wonderful as the beach in Wei Hai, a fantastic place which is
suitable for talking about the fantastic idea on neutrino
telescopes. I am also grateful to Z. Cao and S. Zhou for many
stimulating discussions on UHE cosmic rays and neutrino astronomy.

\end{document}